\documentclass[fleqn,10pt]{wlscirep}
\usepackage[utf8]{inputenc}
\usepackage{graphicx}
\usepackage[T1]{fontenc}
\title{No Love Among Haters:\\Negative Interactions Reduce Online Hate Community Engagement}
\author[1,*]{Daniel Hickey}
\author[2]{Matheus Schmitz}
\author[3,4,5]{Daniel Fessler}
\author[6,7]{Paul E. Smaldino}
\author[2]{Goran Muri\'c}
\author[2]{Keith Burghardt}
\affil[1]{Oregon State University, Department of Botany and Plant Pathology, Corvallis, OR 97331, USA}
\affil[2]{USC Information Sciences Institute, Marina del Rey, CA 90292, USA}
\affil[3]{Department of Anthropology, University of California, Los Angeles, Los Angeles, CA 90095, USA}
\affil[4]{Bedari Kindness Institute, University of California, Los Angeles, Los Angeles, CA 90095, USA}
\affil[5]{Center for Behavior, Evolution, \& Culture, University of California, Los Angeles, Los Angeles, CA 90095, USA}
\affil[6]{Department of Cognitive and Information Sciences, University of California, Merced, Merced, CA 95343, USA}
\affil[7]{Santa Fe Institute, Santa Fe, NM 87501, USA}
\affil[*]{hickeyda@oregonstate.edu}

\begin{abstract}
While online hate groups pose significant risks to the health of online platforms and safety of marginalized groups, little is known about what causes users to become active in hate groups and the effect of social interactions on furthering their engagement. We address this gap by first developing tools to find hate communities within Reddit, and then augment 11 subreddits extracted with 14 known hateful subreddits (25 in total). Using causal inference methods, we evaluate the effect of replies on engagement in hateful subreddits by comparing users who receive replies to their first comment (the treatment) to equivalent control users who do not. We find users who receive replies are \emph{less} likely to become engaged in hateful subreddits than users who do not, while the opposite effect is observed for a matched sample of similar-sized non-hateful subreddits. Using the Google Perspective API and VADER, we discover that hateful community first-repliers are more toxic, negative, and attack the posters more often than non-hateful first-repliers. In addition, we uncover a negative correlation between engagement and attacks or toxicity of first-repliers. We simulate the cumulative engagement of hateful and non-hateful subreddits under the contra-positive scenario of friendly first-replies, finding that attacks dramatically reduce engagement in hateful subreddits. These results counter-intuitively imply that, although under-moderated communities allow hate to fester, the resulting environment is such that direct social interaction does not encourage further participation, thus endogenously constraining the harmful role that these communities could play as recruitment venues for antisocial beliefs.
\end{abstract}
\begin{document}

\flushbottom
\maketitle
%
%
\thispagestyle{empty}

\section*{Introduction}
Hate groups have been shown to cause harm in online environments \cite{Chan2016}, and hate that is spread online can influence offline events \cite{lewis2019online, garland2022impact}. With the rapid growth of the internet, social media platforms such as Reddit\footnote{\url{www.reddit.com}} make it easier than ever for hateful individuals to congregate \cite{caiani2015transnationalization}. The design of Reddit is especially conducive to this phenomenon, as communities, called ``subreddits,'' can be formed to discuss virtually any topic, and participants can hide behind anonymous handles. Although there has been significant research into the spread of online hate \cite{Chandrasekharan2017,Schmitz2022,Chan2016,rajadesingan2020quick} and the spillover of hate between social media communities \cite{Schmitz2022,rajadesingan2020quick}, little is known about what causes users to become active members of hateful communities online. Given that many extremist organizations effectively employ the internet to recruit individuals for violent causes \cite{winter2020online}, it is vital to understand interactions between new contributors and existing participants in hateful subreddits, as this can reveal whether subreddits constitute fora in which indoctrination and incorporation are likely to occur. Previous research indicates that when users interact with newcomers in online communities, those newcomers are more likely to stay \cite{kraut2012building, burke2009feed, burke2011plugged}, especially if the interaction is a positive one \cite{arguello2006talk}. However, it is not known whether this is true of online hate groups, nor whether such groups, being characterized by selective antagonism, are welcoming towards newcomers.

We address this knowledge gap with a causal model-based study to analyze the effect of interactions on hateful and non-hateful subreddits. After utilizing a novel method to extract hateful subreddits from data, we study how interactions on hateful and non-hateful subreddits affect the probability that a user continues to be an active member. We use replies to a user's first post as a treatment, and on each subreddit compare users who received a reply to similar users who posted but never received a reply. We find that, on non-hateful subreddits, new users who received a reply to their first post show a greater likelihood of continuing to be active, yet the reverse is true on most hateful subreddits. We explore why this may be the case using Google's Perspective API \cite{Jigsaw2017} and VADER \cite{hutto2014vader}. We find that, even after controlling for their greater hate speech usage, hateful subreddits show more attacks, toxicity, and negativity, features negatively correlated with the probability a user will continue to post. We use these findings to create a model simulating the contrapositive wherein first-repliers do not reply to first-posters with an attack, toxicity, or negativity, finding that removing hostile language significantly increases member activity in hateful subreddits; because non-hateful subreddits have lower levels of each, their activity is comparatively unchanged. We conclude that although under-moderated hateful subreddits allow hate speech to thrive, the hostility characteristic of these communities discourages newcomers from becoming active members, thereby conceivably limiting the potential impact of these groups.

In summary, our contributions are as follows:
\begin{itemize}
    \item We develop a novel technique to detect hateful subreddits.
    \item We apply causal modeling techniques to determine how a user's first interactions affect their subsequent activity in hateful and non-hateful subreddits.
    \item We quantify how attacks, toxicity, and negative sentiment all contribute to reductions in user activity, and show that these features are more apparent in hateful subreddits.
    \item We simulate how users would behave with more positive interactions, and show that hateful subreddits lose a significant proportion of users to replier vitriol, whereas non-hateful subreddits lack vitriol and are therefore broadly unaffected.
\end{itemize}



\section*{Related Work}
\subsection*{Engagement in Online Communities}

Users have many motivations for joining online communities: exchanging information, acquiring social support, or merely alleviating boredom \cite{ridings2004virtual, khan2017social, lewis2018understanding, yao2021join, stockdale2020bored}, and they may be recruited by friends or acquaintances (a more effective pathway than impersonal advertising) \cite{kraut2012building}. However, not all motivations for joining a community are positive. Users from one community may invade another community with negative posts \cite{jhaver2018online}, which can reduce the overall activity in the invaded community \cite{kumar2018community}. And some communities are, at their core, focused on discussions of antagonistic and hateful aims, including those that are racist, sexist, or xenophobic.  

After a user initially joins a community, various factors contribute to whether they continue participating. When newcomers receive replies to their first posts, they are more likely to keep posting in that community \cite{arguello2006talk, burke2009feed, burke2011plugged, kraut2012building}. The positive impact of the reply is strengthened when it contains positive \cite{arguello2006talk} or personalized \cite{kraut2012building} language. Similarly, new Wikipedia editors who have their first edits reverted are less likely to continue making edits  \cite{halfaker2011don, zhang2006intrinsic}, and Wikipedia therefore encourages users to be gentle with newcomers \cite{wikibite}. However, not all newcomers may be desired by or beneficial for a given community \cite{kraut2012building, choi2008matching}. For example, members of pro-anorexia communities employ gatekeeping tactics to exclude newcomers labeled as ``wannarexics" \cite{boero2012pro}. Similar negative interactions with new users are also seen in Stack Exchange Q\&A boards \cite{Santos2020}. Beyond direct social interaction, users' willingness to adapt to linguistic norms of a community are predictive of how long they will stay in a community \cite{danescu2013no}. Although many users of online health communities stay past their initial motivations for joining (e.g., being diagnosed with cancer), a commonly cited reason for leaving is negative emotions expressed by other users \cite{yao2021join}.

\subsection*{Online Hate Groups}

Members of online hate groups can coordinate in various ways to build stronger coalitions of hate. Users who post toxic material and peddle extremist sources have greater reach than users whose language is positive \cite{phadke2021educators}. Because online hate communities occur on a variety of social media platforms having different moderation policies and design elements, research must situate each platform in the larger online hate ecosystem. Velasquez et al. \cite{velasquez2021online} analyze six online platforms to characterize how hateful content can spread between platforms. Another study analyzing multiple platforms shows how moderating hateful content on one platform can cause hateful content to spread more quickly on other platforms, making a case for the cross-platform regulation of social media \cite{johnson2019hidden}. Facebook is apparently used more by hate groups for actively recruiting members, while Twitter is used to amplify the message of the group and reach a larger audience \cite{phadke2020many}. Research on Reddit reveals their upvote feature may have aided in platforming hateful content on r/The\_Donald \cite{gaudette2021upvoting}. Other studies compare hate speech usage across different social media platforms \cite{zahrah2022comparison, rieger2021assessing, zannettou2018gab}, or evaluate the impact of time spent on social media platforms on hate speech \cite{gallacher2021hate}. Here, we provide greater insight into recruitment dynamics of hate groups on Reddit, further informing how Reddit may compare to other platforms.

Causal inference methods have been applied to further understand hate groups on social media  \cite{founta2021survey}. Matching approaches are common, and have been used to examine the effects of moderation policies \cite{Chandrasekharan2017, chandrasekharan2020quarantined, horta2021platform} and the effects of hateful behavior on other platform members \cite{Schmitz2022, saha2019prevalence}. However, these studies have either examined hate speech outside of hate groups or have been case studies of a small number (two to four) of hateful communities. We expand on such research by making causal inferences in a much larger sample of hateful communities.

\subsection*{Discourse Analysis}

Though interactions with newcomers on online platforms are generally associated with increased engagement by newcomers, the character of interactions is important in this regard \cite{arguello2006talk}. Below we review methods for determining the character of social media text.

Sentiment analysis attempts to identify in text opinions toward a target \cite{medhat2014sentiment}. Many sentiment analysis tools predict ``valence,'' the degree of positivity or negativity in the text. Such tools have been designed using machine learning \cite{zhang2018deep} or lexicon-based approaches \cite{bonta2019comprehensive}. While lexicon-based approaches typically account for fewer lexical features than machine learning approaches, they are more explainable \cite{hutto2014vader}. VADER \cite{hutto2014vader} is a lexicon-based sentiment analysis tool built for social media text that improves performance through additional rules in its design. It has been shown to perform better than other commonly used lexicon-based sentiment analysis tools on social media text \cite{hutto2014vader, bonta2019comprehensive}.

The prevalence of toxic posts has serious implications for the health of online communities. Toxic post detection is thus often used to guide online content moderation \cite{risch2020toxic}. Hate speech and online harassment are often captured within the definition of toxicity, and toxic post detection tools have been shown to perform well on hate detection tasks \cite{van2018challenges,mittos2020and}. A common tool for toxic post detection is the Perspective API, trained on millions of online posts, from diverse sources, the toxicity of each having been annotated by multiple human raters \cite{Jigsaw2017}.

A less widely studied and perhaps more challenging task is to detect personal attacks in online comments. Wulczyn, Thain, and Dixon \cite{wulczyn2017ex} built a classifier to identify these in Wikipedia comments. The Perspective API also includes a tool, trained on New York Times comments, for detecting attacks on other commenters \cite{Jigsaw2017}. It has been used in multiple studies, in conjunction with Perspective's toxicity model and a hate lexicon, to identify directed hate speech \cite{elsherief2018hate, elsherief2018peer, lima2020characterizing}. Other Perspective attributes trained on New York Times data have been used successfully on Reddit \cite{mittos2020and}.

\section*{Methods}

\subsection*{Measuring Engagement}
Within each subreddit, we identified users whose first post was a submission (post that initiates a thread) or a comment (post within a submission) made before 2022, and created the Engagement Risk Ratio ($ERR$) to measure the effect of a reply within a given subreddit. The $ERR$ is defined as the proportion of users who received replies and \emph{continue to engage} (continue to post comments or submissions in another thread) divided by the proportion of users who did not receive replies and continue to engage: 
\begin{equation}
    {ERR} = \frac{\mbox{fraction with replies who continue to engage}} {\mbox{fraction without replies who continue to engage}}
\end{equation}
This is analogous to risk ratios commonly used in, for example, epidemiology \cite{cipriani_nosè_barbui_2007}. Replies are associated with increased engagement if the $ERR$ is greater than one, and decreased engagement if it is less than one.


\subsection*{Detection of Hate Communities}

Defining a hate group as ``an organization or collection of individuals that...has beliefs or practices that attack or malign an entire class of people, typically for their immutable characteristics,''\footnote{\url{https://www.splcenter.org/20200318/frequently-asked-questions-about-hate-groups\#hate\%20group}} we sought to extract subreddits that disparage a class of individuals for properties they cannot change. To quantify this definition, we defined hateful subreddits as those that contain a substantial percent of hate words compared to baseline subreddits. We audited this definition by having auditors extract probable hateful subreddits and compare the contents therein with that of known hateful subreddits.

We began with an initial sample of four hateful subreddits using those examined in prior research on hate communities \cite{Schmitz2022,Chandrasekharan2017}. To this sample we added banned or quarantined subreddits described in the ``controversial Reddit communities'' Wikipedia page \cite{enwiki:1113751282} with strong references to hate in descriptions of their content or reasons for removal; there were 10 such subreddits.

Next, we developed a method for identifying additional hateful subreddits. As members of previously studied hateful subreddits are likely to be members of other hateful subreddits, we sought to retrieve subreddits commonly used by members of the former. We first crawled the posting history of the users in each of the four hateful subreddits examined in Schmitz et al. \cite{Schmitz2022}, from which we obtained a list of subreddits in which each user has been active. For each of those hateful subreddits, we obtained the 1,000 subreddits with the highest proportion of users from the hateful subreddit, where we gathered absolute membership from each subreddit to get the proportion. As there is overlap between the top 1,000 from each hateful subreddit, this resulted in a total of 2,177 subreddits.

We checked each subreddit to see if it is banned, quarantined, or private by checking if it is accessible through the official Reddit API. For each banned, quarantined, or private subreddit with more than 3,000 users, we obtained a list of 100 candidate hate words using SAGE \cite{eisenstein2011sparse}, following Schmitz et al. \cite{Schmitz2022}. SAGE compares a target corpus to a baseline corpus to find the words most characteristic of the former. For our baseline corpus, we used a sample of randomly selected Reddit posts using the Reddit API. After the top 100 most characteristic words of each subreddit were obtained, three human annotators (who are also authors of this paper) rated each word as 0 $=$ not hateful, 1 $=$ sometimes hateful, or 2 $=$ always hateful (the context of the subreddits were taken into account when annotating). The annotations resulted in a Fleiss Kappa score of 0.5, indicating moderate agreement among raters. Words with a total score of four or greater were classified as hateful. Subreddits with greater than five hate words were classified as hateful; 14 subreddits met this criterion. Three of the subreddits (r/TheRedPill, r/milliondollarextreme, and r/CringeAnarchy) were independently obtained from Wikipedia's (non-exhaustive) list of highly prominent hateful subreddits, supporting the validity of our method. Table~\ref{tab:subreddits_list} displays the full list of hateful subreddits. All these hateful subreddits contain data on both comments and submission except for four: GreatApes, CoonTown, NeoF*g, and FatPeopleHate. We do not believe this significantly affects our results. Automated accounts were removed both as newcomers and repliers by excising accounts with usernames matching certain substrings, such as ``bot'', and manually analysing the remaining accounts with highest activity, following Schmitz et al. \cite{Schmitz2022}. As automated accounts make up a small proportion of accounts in each subreddit, this step does not notably impact our overall results.

\begin{table*}[h]
    \centering
    \begin{tabular}{ccccc}
    \textbf{Subreddit}   & \textbf{Num. Users} & \textbf{Type of Hate} & \textbf{Source} & \textbf{Similar size non-hateful subreddit} \\ \hline
    AskTRP               & 87,367                   & Sexist                & Banned list     & AnimeSketch                            \\
    Braincels            & 45,817                   & Sexist                & Wikipedia       & OrnaRPG                           \\
    Incels               & 37,619                   & Sexist                & Prior research  & samuraijack                                     \\
    MGTOW                & 125,481                   & Sexist                & Wikipedia       & ApplyingToCollege                                   \\
    TheRedPill           & 126,648                  & Sexist                & Wikipedia     & ask                          \\
    TruFemcels           & 12,447                   & Sexist                & Wikipedia       & Fallout76MarketPlace                                  \\ \hline
    CoonTown             & 11,042                   & Racist                & Prior research  & MCPE                                 \\
    GreatApes            & 3,139                    & Racist                & Prior research  & sneakerreps                              \\
    WhiteRights          & 9,191                    & Racist                & Banned list     & ForeverAloneWomen                                        \\
    CCJ2                 & 4,424                    & Racist                & Banned list     & prozac                              \\ \hline
    Honkler              & 3,086                    & Alt-right             & Wikipedia       & PixelGun                              \\
    frenWorld            & 17,604                   & Alt-right             & Wikipedia       & Pikabu                              \\
    SJWHate              & 29,384                   & Alt-right             & Banned list     & Paladins                               \\
    milliondollarextreme & 26,563                   & Alt-right             & Wikipedia     & RocketLeagueExchange                               \\ \hline
    CringeAnarchy        & 246,587                  & Trolling/harassment   & Wikipedia     & NoFap                             \\
    trolling             & 4,800                    & Trolling/harassment   & Banned list     & DannyGonzales                                     \\
    FuckYou              & 12,803                    & Trolling/harassment   & Banned list     & TimPool                                     \\
    Neof*g               & 945                      & Trolling/harassment   & Wikipedia       & dragonvale                          \\ \hline
    TrollGC              & 5,807                    & Anti-LGBT             & Banned list     & ninjavoltage                               \\
    GenderCritical       & 55,004                   & Anti-LGBT             & Wikipedia       & truerateme                         \\
    dolan                & 11,679                    & Anti-LGBT             & Banned list     & Soundbars                                    \\ \hline
    ImGoingToHellForThis & 450,915                  & General hate          & Banned list     & Drugs                                     \\
    opieandanthony       & 31,793                   & General hate          & Banned list     & DragonballLegends                                         \\
    DelrayMisfits        & 4,381                    & General hate          & Banned list     & newsokunomoral                                    \\ \hline
    FatPeopleHate        & 56,340                   & Fat-shaming           & Prior research  & Planetside                              
    \end{tabular}
    \caption{Hateful subreddits used in analysis. The subreddits capture a variety of sizes and targeted groups.}
    \label{tab:subreddits_list}
\end{table*}

\subsection*{Matching}
To make causal inferences about hateful subreddits and their users, we employed a Mahalanobis distance matching approach \cite{Stuart2010} wherein we compare similar users in each community, one of whom receives a reply. We explore two related analyses: first, estimating the effect of a reply within each subreddit. Second, estimating the overall effects of replies based on subreddit type. For each inference, we perform a separate user matching. Our matching process derives from methods used by Schmitz et al. \cite{Schmitz2022}. Features were measured prior to the treatment. In the matching process, a user in the treatment group is chosen, the most similar user in the control group is selected as its pair, and the process is repeated until there are no unmatched users in the treatment group.

To estimate the direct effect of replies within each subreddit, we matched users who received replies to similar users who did not receive replies, based on four potential confounders: 
the time elapsed between account creation and their first post in the studied subreddit;
the nest level (with 1 indicating a top-level comment, 2 indicating a direct reply, 3 indicating a reply to that reply, etc.) of their first comment; 
the sentiment valence of their first post, as measured by VADER \cite{hutto2014vader}; 
and the number of words in their first post. 
Features like these have previously been shown to impact users' likelihood of receiving replies in online discussions \cite{arguello2006talk}. This process was applied to both hateful and non-hateful subreddits. We measure the effects of replies to submissions and replies to comments separately due to the qualitative differences between each type of post - therefore, users who join subreddits by making submissions are in a separate matching pool from users who join subreddits with comments. Samples of users in each matching pool were limited to 30,000 per subreddit. 

To compare the overall effects of replies in hateful subreddits to the effects in non-hateful subreddits, we collected 324 random subreddits by obtaining random posts from the Reddit API, identifying the subreddit associated with each post, and using the Pushshift API \cite{Baumgartner2020} to crawl all subreddits with greater than 10,000 members and less than 2 million members. Larger subreddits were excluded due to the time required to crawl subreddits using Pushshift. We used Mahalanobis distance matching to pair each hateful subreddit with a non-hateful subreddit from this sample. Each matched non-hateful subreddit was manually inspected to ensure the posts were not primarily made by automated accounts.

Comparing the engagement in banned (hate) subreddits and non-banned subreddits can bias the results of our analysis because 
posts by users who would otherwise be active after the moment a subreddit is banned would not appear in the dataset. In this case, the engagement of users might appear lower simply because the posts they would have made could not be created. We account for this in the process of matching subreddits. 
Namely, for each hateful subreddit, we obtain the 90th percentile of the time taken for newcomers to post a second time.
 This value, alongside the subreddit size, is used for matching each hateful subreddit with a non-hate one. Limiting to the 90th percentile prevents our results from being skewed by users who could not post a second time because the subreddit had been banned.
Moving forward, a ban was simulated for each non-hateful subreddit by excluding posts made after their respective matched hateful subreddit's ban date. Additionally, users who posted after the 90\% percentile return time before the ban were excluded, as they may have been unable to keep posting due to the (simulated) ban.

\subsection*{Analyzing Content of Replies}

To identify replies that may discourage users from continuing to engage within subreddits, we used the Perspective API, a collection of models used for the detection of toxic online posts \cite{Jigsaw2017}. The Perspective API has been widely utilized and validated in a variety of domains, \cite{pavlopoulos2019convai, frimer2022incivility, saveski2021structure}, including Reddit \cite{kumar2022understanding,mittos2020and,rajadesingan2020quick}. We measured two attributes from the Perspective API to better understand the content of replies: ``toxicity'' and ``attack on commenter.'' Although we analyze both subreddit comments and submissions, we employ the Perspective terms, labeling the metric ``attack on commenter'' to both comments and submissions for terminological consistency. Perspective API defines a toxic post as ``a rude, disrespectful, or unreasonable comment that is likely to make people leave a discussion.'' The attack on commenter attribute is a metric trained using New York Times comments. The attribute is meant to identify hurtful posts directed at other members of a given comment thread. The outputs from the toxicity and attack on commenter models are probabilities that the posts are toxic or attacks, respectively. The Pearson correlation between these attributes is relatively weak ($r=0.012$, calculated from replies in all hateful subreddits), hence these constitute independent metrics. We also explored the severe toxicity metric from Perspective by replacing regular toxicity with severe toxicity in our analysis and found the overall results to be similar. We use regular toxicity only because it has been much more thoroughly validated across a range of datasets \cite{pavlopoulos2019convai, frimer2022incivility, saveski2021structure}, including Reddit \cite{kumar2022understanding,mittos2020and,rajadesingan2020quick}. Perspective also recommends a probability threshold of 0.7 for their toxicity metric, which we do not perform in our main analyses, instead treating the metric as a continuous variable. However, we repeated our analyses using a threshold of 0.7 for both toxicity and attack on commenter metrics and found similar results. In addition to the attributes from the Perspective API, we measured the sentiment of each reply to a first post using VADER \cite{hutto2014vader}, a lexicon and rule-based sentiment analysis tool specifically geared towards social media text. Whereas the attributes from the Perspective API are intended to capture negative or hurtful replies, sentiment analysis allows us to capture the valence of replies. 

To understand how strongly these models are affected by hateful language, we tested how sensitive the outputs of each model are in response to hate words. We built a lexicon of 260 hate words from all 25 subreddits using the process of obtaining and annotating words described in section 3.2. We then calculated the attack on commenter, toxicity, and sentiment probabilities for each post with hate words and each post with hate words replaced with neutral synonyms (for example the n-word was replaced with ``black person''), see hate speech lexicons and replaced words here (WARNING: Contains offensive terms): \url{https://anonymous.4open.science/r/reddit_lexicons-8628/}. We ensured the neutral word replacements did not belong to the VADER lexicon, as that could bias our results. The average toxicity, attack on commenter, and sentiment outputs for replies in each subreddit were then compared, including replies that did not originally contain hate words. 

\subsection*{Simulating Growth of Subreddits}

\begin{figure}[h]
    \centering
    \includegraphics[width=0.75\columnwidth]{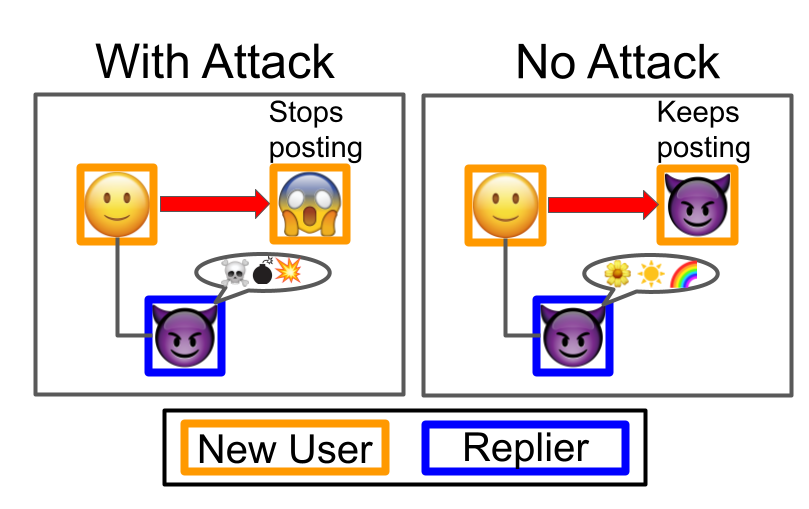}
    \caption{Schematic of hateful subreddit growth simulation. A mixed effect logistic regression model predicts whether a user continues posting or leaves the subreddit. Predictions are counted to calculate the cumulative number of engaged users in a subreddit.}
    \label{schematic}
\end{figure}

To understand how replies and their content impact the growth of hateful subreddits, we created a mixed effect logistic regression model to measure the likelihood a user continues to post after their first post (what we term an ``engaged'' user). We feed in the subreddit as random effects and use the features of the first reply to a user's post on that subreddit (the reply's Perspective API attribute probabilities and VADER sentiment scores) as fixed effects. We fit separate models for comments and submissions, as well as separate models for hateful and non-hateful subreddits. We used these to measure the probability a user will be engaged for all posts (we used the same dataset for training and testing because our goal is modeling the dataset being studied, rather than building a system for out-of-sample prediction). To ensure our models are valid, we computed the variance inflation factor (VIF) for each one, finding that no coefficient reached a VIF greater than 2.5, which is consistent with non-multicollinear features \cite{senaviratna2019diagnosing}. 
We then measured the probability each first-poster would continue, given their first replies, and simulated a user as either continuing or not depending on whether a Bernoulli random variable of this probability was 1 or 0, respectively. 
We created two simulations from these models, shown in Fig.~\ref{schematic}, to test how user engagement would change under ``nicer'' subreddits. In one, we used the true features from the replies the users received to predict whether each user is engaged. In the other, we set the attack on commenter and toxicity probabilities of all replies to zero and set the sentiment scores of negative sentiment replies to zero, then made the same predictions. For each subreddit, we added up the total number of engaged users in each scenario and calculated the percent increase in engaged users for the scenario with no toxicity, negativity, or attack relative to the default scenario. These simulations include users who make comments as their first post as well as users who make submissions as their first post.

\section*{Results}

\begin{figure}[h]
    \centering
    \includegraphics[width=0.9\columnwidth]{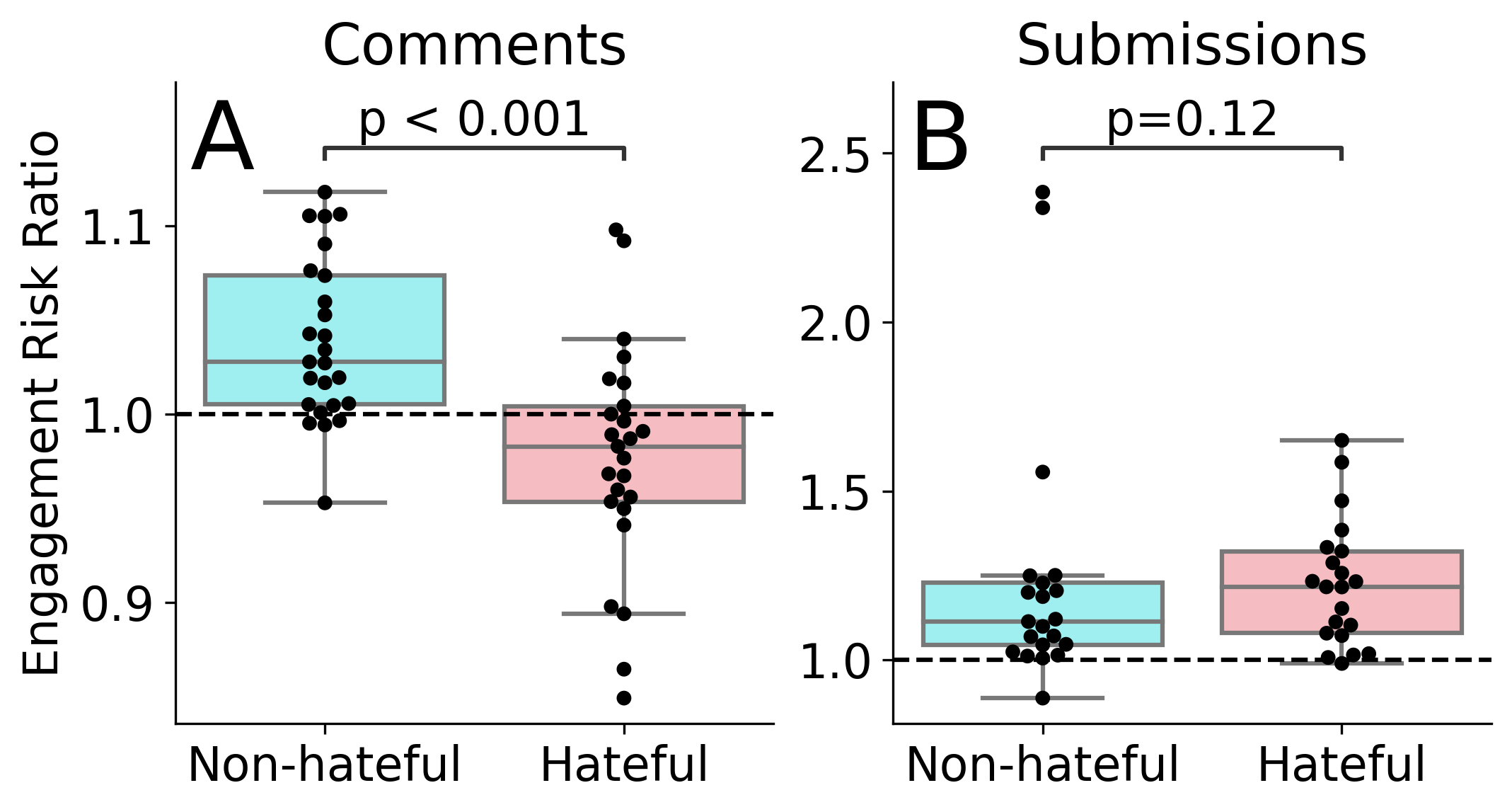}
    \caption{Replies to comments in hateful subreddits lead to significantly less engagement than replies to comments in non-hateful subreddits. Distributions of engagement risk ratios for different subreddit types, separated by users who make comments as their first post (A) and users who make submissions as their first post (B). Points represent engagement risk ratios of individual subreddits, while boxplots summarize the overall distributions for each subreddit type. Values of engagement risk ratios greater than one imply replies are associated with more active users, while values less than one imply replies are associated with less active users. The lines in the boxes represent medians, while boxes and outer lines represent the inter-quartile range and 95\% quantiles, respectively.}
    \label{ERR_distribution}
\end{figure}

\textbf{Replies to comments in hate groups lead to lower engagement.} The vast majority (78\%) of all first-posts within our dataset are comments. Among these comments, hateful subreddits have a significantly lower $ERR$ than non-hateful subreddits (Fig~\ref{ERR_distribution}). We use a Wilcoxon signed-rank test to assess significance ($T=44, p<0.001$) \cite{Wilcoxon1945}. The mean $ERR$ for hateful subreddits is below one (0.98), while the mean $ERR$ for non-hateful subreddits is above one (1.03), indicating replies have opposite effects, on average, for each subreddit type. 
For users whose first post is a submission, there is no significant difference in $ERR$ between the two subreddit types. The mean $ERR$ is above one for both types, meaning these first-posters are generally encouraged to keep posting after receiving replies.

\begin{figure*}[h]
    \centering
    \includegraphics[width=0.85\textwidth]{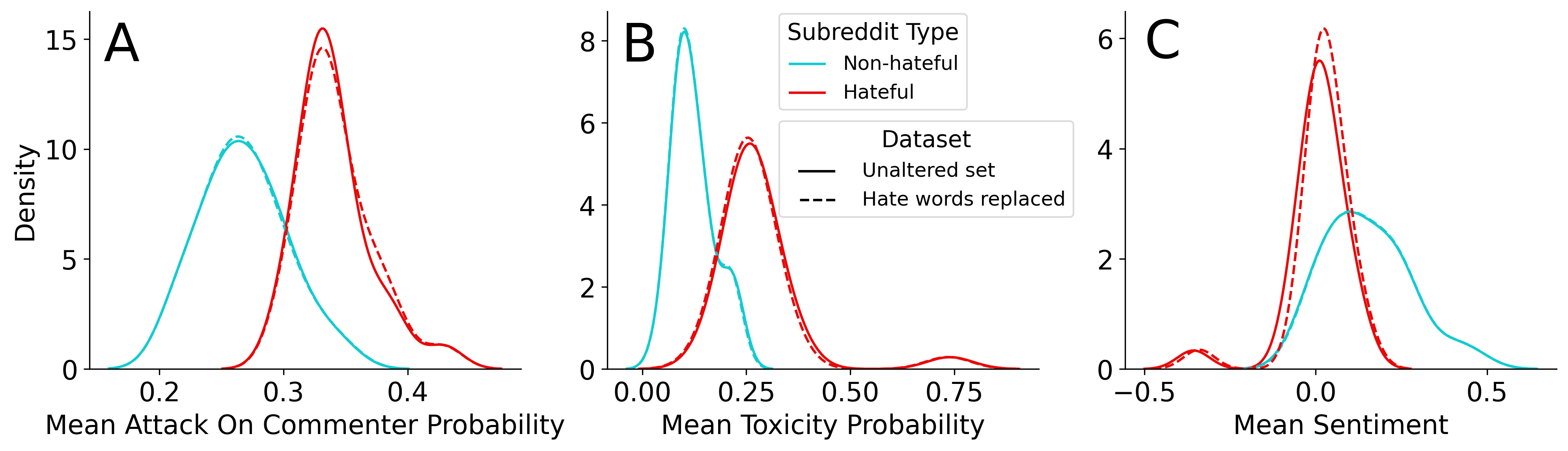}
    \caption{Attacks and toxicity are higher in hateful subreddits replies, while sentiment is lower. Distributions across all users for (A) attack on commenter probability, (B) toxicity probability, and (C) valence for replies to comments in hateful and non-hateful subreddits. Substitution of synonymous non-hate terms does not alter the ability of the models to distinguish between subreddit types. Results are similar when analyzing replies to submissions.}
    \label{reply_content_distributions}
\end{figure*}

\textbf{Attacks, toxicity, and negativity are prevalent in the replies of hateful subreddits.} Replies in hateful subreddits are generally more toxic, negative, and contain more attacks than non-hateful subreddits. Fig.~\ref{reply_content_distributions} depicts differences in (A) mean attack on commenter probabilities, (B) mean toxicity probabilities, and (C) mean sentiment for replies to comments. These results are similar for replies to submissions. 
A Wilcoxon signed-rank test  \cite{Wilcoxon1945} reveals the mean attack on commenter probability for replies to comments is significantly higher in hateful subreddits than non-hateful subreddits ($T=1$, $p < 0.001$). The mean toxicity probability is also higher ($T=0$, $p < 0.001$), while mean sentiment is lower ($T=0$, $p < 0.001$). For replies to submissions, the mean attack on commenter probability is higher in hateful subreddits, though the difference is not significant. However, the mean toxicity remains significantly higher in hateful subreddits ($T=1$, $p < 0.001$) and the mean sentiment is significantly lower ($T=25$, $p < 0.001$).
Additionally, for hateful subreddits, there are significant differences in toxicity between replies to submissions and replies to comments. The mean toxicity probability is greater in replies to comments compared to replies to submissions ($T=55$, $p=0.04$). Other differences are not significant. 
When assessing the impact of hate words on attacks, toxicity, and sentiment, replacing hate words with neutral words does not alter their distributions nearly enough to account for the differences between hateful and non-hateful subreddits. The reason for this small difference is because posts with hate words are in the minority of all subreddits, despite their impact on toxicity \cite{van2018challenges}.

\textbf{Content of replies are predictive of whether users will leave a subreddit.} To understand how the content of replies relates to subreddit engagement, we built mixed effect logistic regression models, where the response variable was whether or not a user continues to post given they either made a submission or comment as their first post, and the independent variables were whether or not they received a reply, and, if so, the sentiment, toxicity, and attack on commenter scores of that reply. Table~\ref{mixed_effect_model} summarizes the results. 

The models show that, for new users, each reply received increases their likelihood of remaining engaged in that subreddit, after controlling for covariates. The models also show a much stronger negative effect of attacks on posters than of toxicity or sentiment. The models suggest that the higher attack on commenter probabilities in hateful subreddits are a principal contributing factor to the lower $ERR$s in these subreddits. We find a non-significant relationship between engagement and toxicity or sentiment in non-hateful subreddits, and that, while toxicity appears to reduce the likelihood of engagement in comments within hateful subreddits, it increases engagement for submissions. Finally, while positive sentiment increases engagement, especially for commenters in hateful subreddits, this effect is not significant for submissions, possibly because the sentiment is directed at the content rather than the user, therefore users are less impacted.

\begin{table*}[h]
\centering
\begin{tabular}{lcc|cc}
\multicolumn{1}{c}{}                                                       & \multicolumn{2}{c}{\textbf{Comments}}                   & \multicolumn{2}{c}{\textbf{Submissions}}                  \\
\multicolumn{1}{c}{\textbf{}}                                              & Hate                   & Non-hate                 & Hate                     & Non-hate                 \\ \hline
\textbf{Fixed Effects: Coefficient $\pm$ Standard Error} & & &                                                                \\
Mean Intercept                                                             & \textbf{0.085 $\pm$ 0.07} & 0.13 $\pm$ 0.069            & \textbf{-0.435 $\pm$ 0.101} & -0.155 $\pm$ 0.126          \\
\sc{Reply}                                                & \textbf{0.09 $\pm$ 0.01}  & \textbf{0.104 $\pm$ 0.012}  & \textbf{0.366 $\pm$ 0.020}  & \textbf{0.228 $\pm$ 0.021}  \\
\sc{Reply} $\times$ \sc{Sentiment}       & \textbf{0.11 $\pm$ 0.01}  & \textbf{0.060 $\pm$ 0.013}  & -0.015 $\pm$ 0.022          & \textbf{0.060 $\pm$ 0.025}  \\
\sc{Reply} $\times$ \sc{Toxicity}        & \textbf{-0.05 $\pm$ 0.02} & 0.018 $\pm$ 0.03            & \textbf{0.161 $\pm$ 0.037}  & 0.038 $\pm$ 0.061           \\
\sc{Reply} $\times$ \sc{AttackOnCommenter} & \textbf{-0.35 $\pm$ 0.02} & \textbf{-0.168 $\pm$ 0.022} & \textbf{-0.255 $\pm$ 0.037} & \textbf{-0.206 $\pm$ 0.039} \\ \hline
\textbf{Random Effects: Variance}    & & &                                                                              \\
Subreddit                                                                  & 0.16                      & 0.15                        & 0.24                        & 0.38                       
\end{tabular}
    \caption{Logistic mixed effect model for the probability a user will continue to be active after their first post. Coefficients in bold indicate $p < 0.05$.}
    \label{mixed_effect_model}
\end{table*}

\begin{figure*}[h]
    \centering
    \includegraphics[width=0.85\textwidth]{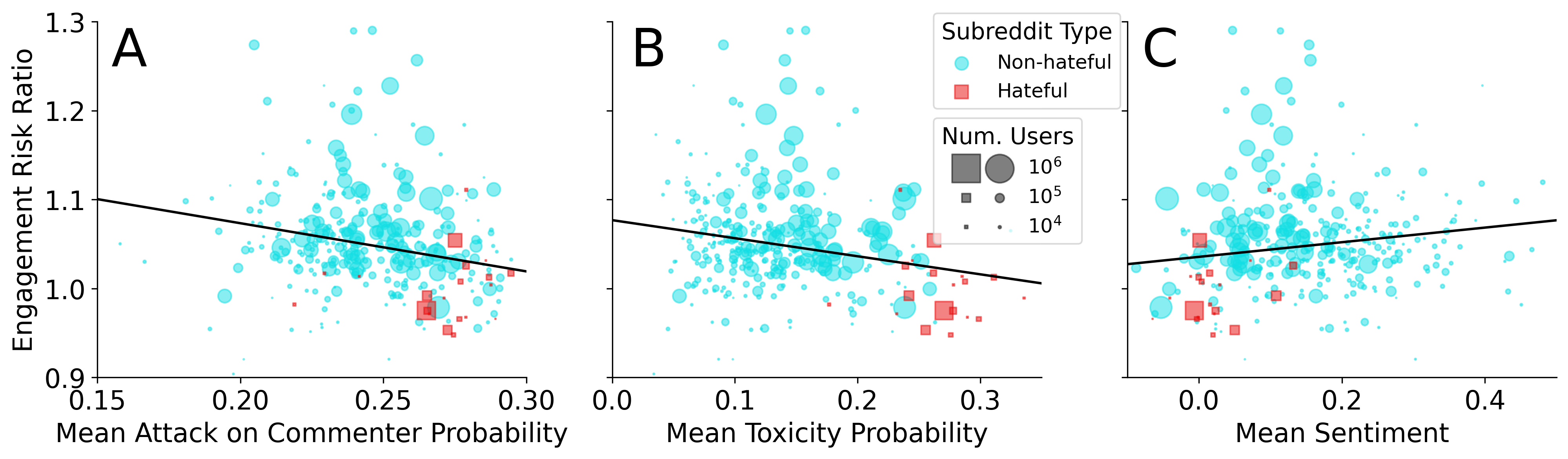}
    \caption{Engagement risk ratios of comments are correlated with attacks, toxicity, and sentiment. Relationships between engagement risk ratios and mean attack on commenter probability (A), toxicity (B), and sentiment (C) of subreddits.}
    \label{correlation_figure}
\end{figure*}

\textbf{Engagement risk ratios of replies to comments in subreddits are correlated with reply attributes.} Figure \ref{correlation_figure} shows a negative correlation between the $ERR$s for newcomers who make comments as their first post of subreddits and the mean toxicity of replies they receive (Spearman correlation coefficient, $r=-0.21, p=0.001$), as well as the mean attack on commenter probabilities ($r=-0.24, p < 0.001$). There is also a positive correlation between mean sentiment of replies and $ERR$s ($r=0.17$, $p=0.001$). This 
indicates that, generally speaking, subreddits where negative, toxic, or attacking language are prevalent lead to less engagement through direct social interactions.

\begin{figure*}[h]
    \centering
    \includegraphics[width=0.8\textwidth]{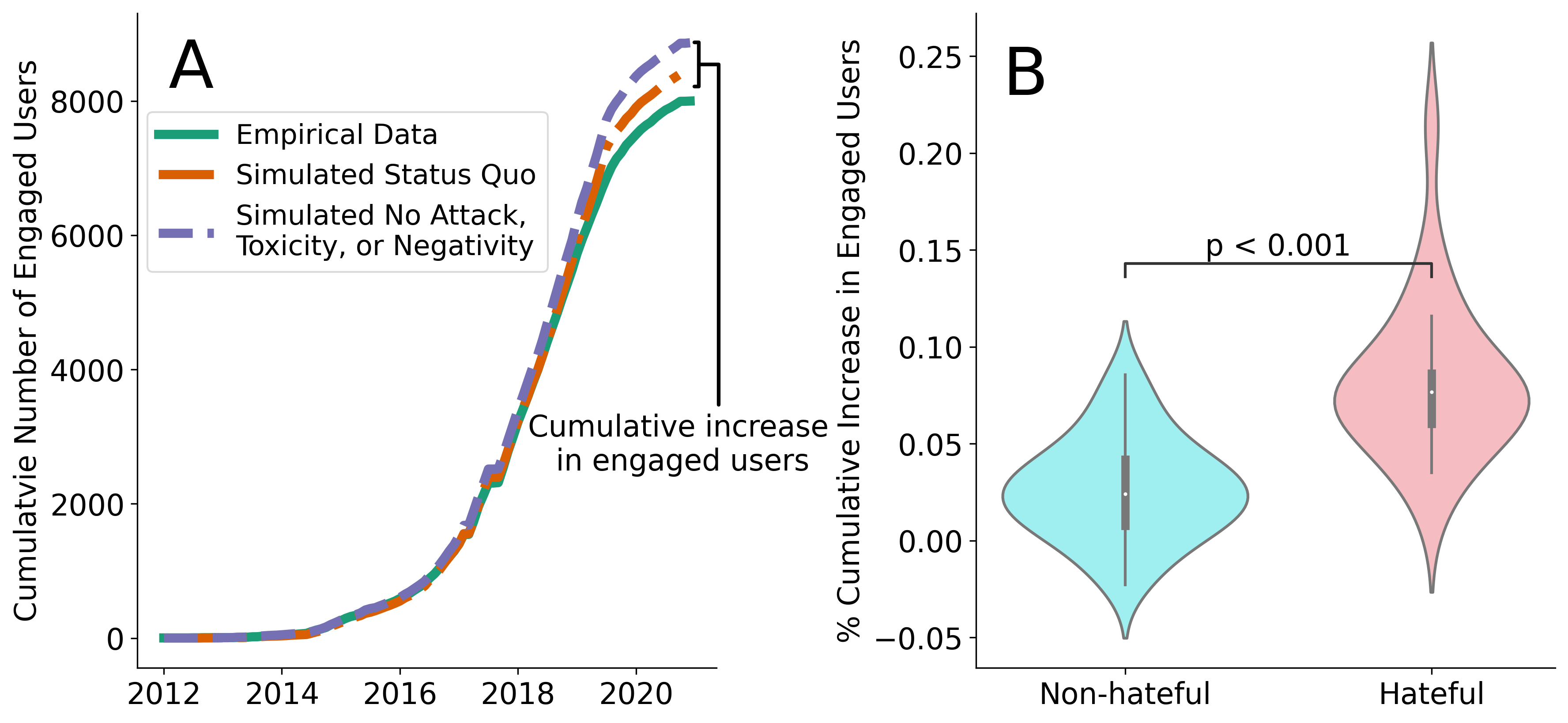}
    \caption{Simulated growth of subreddits. (A) Empirical and simulated growth curve of engaged users in r/MGTOW. (B) Predicted increases in engaged users for hateful and non-hateful subreddits given attack on commenter probability, toxicity, and sentiment of negative replies are set to zero. The white dots represent medians, while the thick gray bars represent the inter-quartile range, and the thin gray lines represent 95\% quantiles.}
    \label{simulation}
\end{figure*}

\textbf{Simulations show negative impact of negativity, toxicity, and attacks on subreddit growth.} Through our simulations, for both hateful and non-hateful subreddits, we observe higher cumulative numbers of engaged users in the contrapositive scenario where there are no attacks, toxicity, or negativity in replies (Fig~\ref{simulation}). The average relative increase from the simulated status quo is much higher for hateful subreddits, indicating that hateful subreddits grow more slowly than they potentially could due to the nature of their replies.

\section*{Discussion}

We analyzed replies to first posts in a matched sample of 25 hateful and 25 non-hateful subreddits to assess the overall effect that replies have on engagement. We find that replies are more likely to cause users to leave hateful subreddits than to continue engaging with them, while the opposite is true for non-hateful subreddits. It is only after controlling for the content of replies in hateful subreddits that the effect of replies on engagement becomes positive -- in other words, it is the negative, attacking, or toxic nature of replies in hateful subreddits that cause receiving a reply to discourage, rather than encourage, new contributors to continue to participate. Below we discuss the implications of these findings for online policymakers, community moderators, and researchers.

\textbf{Hate groups hate themselves.} A principal finding is that attack on commenter scores are much higher in hateful subreddits than in non-hateful subreddits, especially when the first post is a comment rather than a submission (Fig~\ref{reply_content_distributions}A). Unlike sentiment and toxicity, which can detect negative or toxic comments directed at a particular identity group, the attack on commenter metric is an indicator of rude or disrespectful comments directed at other members of the hate community. Members of hateful subreddits are thus generally more disrespectful towards their own communities than are members of non-hateful subreddits. Members of hateful subreddits are thus \emph{generally} more antisocial, as their animosity is not limited to a single identity group, but also applies to others who hold their same prejudices. This is consistent with prior research revealing high levels of psychopathy among online haters \cite{sorokowski2020online}. 

\textbf{Hate groups are self-defeating.} Most social interactions in hateful subreddits decrease engagement within these communities. In hateful subreddits, we see a high prevalence of the types of replies to comments predictive of users leaving subreddits, indicating that users are not regulating their behavior in ways that would encourage newcomers to remain. It is therefore possible that, due to the nature of social interactions within their communities, hate groups on Reddit pose less of a threat in terms of recruitment and radicalization than is true of other online platforms. This finding also makes the case that it is in a platform's best interests to employ some level of moderation in order to maximize its adoption and growth. Quite simply, creating a lawless land is bad for business. That said, replies to users whose first post is a submission encourage \emph{more} activity compared to posts without a reply, possibly because such users are generally more eager to engage in the community regardless of reply toxicity.  

Given the general vitriol of hateful subreddits, the rarity of attacks, toxicity, and negativity within a given online hate community may serve as a useful marker that said community poses a significant threat as a venue for recruitment and radicalization. However, caution is in order when extrapolating from our results. Given the psychological profile of online haters, the hostility characteristic of replies in hateful subreddits could potentially be attractive to other haters. It is therefore possible that, while attacking, toxic, or negative replies reduce new user engagement, they may indirectly \emph{increase} the engagement of other users in the community. Moreover, in the minority of users whose first post is a submission, the toxicity of replies might be directed at whoever is mentioned in the submission rather than the submission's author, thus it could encourage user engagement. 
Additionally, previous research has found that members of hate communities on Facebook who post toxic and inflammatory content are effective at spreading their messaging \cite{phadke2021educators}, and that more toxic comments tend to be more popular on Facebook \cite{kim2021distorting}. Furthermore, there are other ways to measure the success of online communities beyond user retention or number of users \cite{cunha2019all}. Social interactions may therefore affect hateful subreddits in ways not explored in this study.

The attributes of replies have stronger effects in hateful subreddits compared to non-hateful subreddits. That said, a lack of association between toxicity and engagement in non-hateful subreddits may be because the content of replies with high toxicity varies between the different subreddit types. For example, high-toxicity replies in hateful subreddits might be more likely to be insulting, while high-toxicity replies in non-hateful subreddits could be posts with a high prevalence of profanity but no insulting material. 

\subsection*{Limitations and Future Directions}

While we find that replies make newcomers in hateful subreddits less likely to remain active in those communities, there remain additional challenges and questions about recruitment in online hate groups. 

\textbf{Behavior outside of hateful subreddits.} We examined users' likelihood of being engaged in one self-contained community after their first post in said community. However, there are opportunities for them to engage in other hate communities, or exhibit hate behavior beyond the subreddit of interest. Given prior research on how toxicity begets toxicity in online environments \cite{kim2021distorting, cheng2017anyone, saveski2021structure}, it is possible that users who leave hateful subreddits due to toxicity or attacks could be spreading that content elsewhere. Analyzing whether users who leave a hateful subreddit then become engaged in other hateful subreddits, or use hate material to antagonize users in non-hateful subreddits, can illuminate how social interactions in hateful subreddits influence the radicalization of newcomers.

\textbf{More specific content analysis of replies.} Our analysis focused on negative dimensions of text, such as the toxicity or attack on commenter metrics. However, previous research indicates that members of hate groups use particular types of arguments to recruit new members, such as appeals to fear \cite{phadke2020many}. Investigating the presence of such arguments in social interactions in Reddit hate groups and their internal dynamics could clarify the prevalence of these replies and their effects on recipients of such messages.

\textbf{Hate group recruitment outside Reddit.} Previous research suggests that hate groups use different online platforms for different purposes in growing their communities \cite{phadke2020many}, and negative language is more prevalent on certain platforms \cite{zahrah2022comparison}. Replicating our studies on other platforms, such as Facebook, may provide further insight into where online hate group recruitment occurs, as well as illuminate Reddit's role in the larger online hate ecosystem.

\textbf{Opinions of newcomers in hateful subreddits.} Our research shows how antagonism by members of hateful subreddits deters newcomers. However, we do not know between \emph{whom} these antagonistic interactions are occurring. How radicalized are the newcomers who receive antagonistic replies? While we attempted to control for this using causal inference methods, examining who receives antagonistic replies upon joining hateful subreddits could be informative, as it is possible that most newcomers who receive antagonistic replies are ``outsiders'' who enter the community in order to challenge the general views of the hateful subreddit.

\textbf{Validating experimental Perspective API attributes.} Several papers have validated the Toxicity metric of the Perspective API on a range of domains, including Reddit, and our work shows the efficacy of the metric. To further confirm the efficacy of the API, especially on the ``attack on commenter'' metric, which was underexplored in previous analyses, we collected 200 posts at random from hateful subreddits and labeled them as toxic or attacks on commenters. Three of the authors labeled these posts, and, where there were disagreements, the majority decision was used. 
The ROC-AUC scores for the attack on commenter and toxicity metrics are 0.85 and 0.86, respectively, 
thus demonstrating reasonable accuracy. Along with their toxicity model, Perspective provides a ``severe toxicity'' model designed to be less sensitive to hateful words than the toxicity model. Due to the elevated presence of hate terms in our subreddits of interest, we considered using this model, which has an ROC-AUC of 0.84. All results for regular toxicity are similar to those generated using the severe toxicity metric. In the future, a new set of models fine-tuned on Reddit data could create better metrics of toxicity, attacks, and sentiment within Reddit communities.

\textbf{Community-specific discourse analysis.} Content that is generally seen as negative or toxic on social media platforms may not be viewed as such by members of hate groups, as the interpreted meaning of online communication may differ greatly depending on its audience \cite{van2022strategic}. This makes the identification of antagonistic comments within hate groups challenging. While the attack on commenter attribute from the Perspective API appears to effectively identify antagonistic comments within hate groups, it was designed to be used in an entirely different domain. Community-specific tools that address hostile comments directed towards members of hate communities may prove even more accurate, and may provide greater insight into the presence of directed antagonism within hate groups.


\subsection*{Broader perspective, ethics and competing interests}
All data were collected from a public dataset, and all identifiable information was removed prior to analysis. Our findings can contribute to understanding how hate communities form, and why some remain comparatively small. We are cognizant that those who seek to promote prejudice and intolerance could exploit our findings. Nonetheless, we believe that, on balance, it is in society's interests that work such as this be conducted, as shedding light on the dynamics of interactions in contexts in which antisocial views are amplified can guide efforts to create more tolerant online communities, for example by providing tools that can assist platforms in moderating their content.

\bibliography{citations}

\section*{Acknowledgements}

Funding for this work is provided through the USC-ISI Exploratory Research Award, NSF (award \#), and through DARPA (awards \# HR0011260595 and \# HR001121C0169).

\section*{Author contributions statement}
All authors designed research; D.H. and M.S. performed analysis; all authors reviewed the manuscript. 

\section*{Additional information}

\textbf{Competing interests.}  
The authors declare no conflicts of interest.
\end{document}